\documentclass[conference, compsoc]{IEEEtran}
\usepackage{amsmath}
\usepackage{amsthm}
\usepackage{amsfonts}

\usepackage[ruled, linesnumbered, lined]{algorithm2e}

\newtheorem{definition}{Definition}

\newtheorem{lemma}{Lemma}
\newtheorem{theorem}{Theorem}
\newtheorem{proposition}{Proposition}

\def\val{\textsc{val}}
\def\proposed{\textsc{proposed}}
\def\written{\textsc{written}}
\def\wlr{\textsc{writtenOld}}

\def\history{\textsc{history}}
\def\C{\textsc{C}}

\def\proposer{\diamond\textit{-proposer}}
\def\out{\textit{out-connected}}
\def\silent{\diamond\textit{-silent}}
\def\leader{\textit{leader}}
\def\m{m}
\def\h{\textsc{h}}

\def\receive{\textit{receive}}
\def\endofround{\textit{end-of-round}}

\def\initialize{\textit{initialize}}
\def\compute{\textit{compute}}
\def\send{\textit{send}}

\def\delivered{\textsc{delivered}}

\def\get{\textit{get}}
\def\add{\textit{add}}

\def\qtrue{\textbf{true}}
\def\qfalse{\textbf{false}}

\begin{document}

\title{Fault-Tolerant Consensus in Unknown and Anonymous Networks}

\IEEEoverridecommandlockouts

\author{
	\IEEEauthorblockN{Carole Delporte-Gallet, Hugues Fauconnier and Andreas Tielmann}\thanks{Carole Delporte-Gallet and Hugues Fauconnier were supported by grant ANR-08-VERSO-SHAMAN. Andreas Tielmann was supported by grants from R\'egion Ile-de-France.}
	\IEEEauthorblockA{LIAFA, University Paris VII, France \\
						\texttt{\{cd,hf,tielmann\}@liafa.jussieu.fr}
	}
}

\maketitle

\begin{abstract}
This paper investigates under which conditions information can be reliably shared and consensus can be solved in unknown and anonymous message-passing networks that suffer from crash-failures.  We provide algorithms to emulate registers and solve consensus under different synchrony assumptions. For this, we introduce a novel pseudo leader-election approach which allows a leader-based consensus implementation without breaking symmetry. 

\end{abstract}

\section{Introduction}

Most of the algorithms for distributed systems consider that the number of processes in the system is known and every process has a distinct ID. However, in some networks such as in wireless sensors networks, this is not necessarily true. Additionally, such networks are typically not totally synchronous and processes may suffer from failures such as crashes.

Designing protocols for such networks is especially intricate, since a process can never know if its messages have been received by all processes in the system. In this paper, we investigate under which conditions information can be reliably shared and consensus can be solved in such environments.   

Typically, in systems where no hardware registers are available, one makes additional assumptions to be able to reliably share information, e.g.~by assuming a correct majority of processes. However, these techniques assume also some knowledge about the total number of processes. With processes with distinct identities, the requirements to emulate a register have been precisely determined by showing that the quorum failure detector $\Sigma$ is the weakest failure detector to simulate registers in asynchronous message passing systems \cite{200377/IC}. But again, this approach fails due to the lack of identities in our anonymous environment.

To circumvent these problems, we assume that the system is not totally asynchronous, but assume the existence of some partial synchrony. We specify our environments by using the general round-based algorithm framework (GIRAF) of \cite{1449454}. This has two advantages: (i) it is easy to precisely specify an environment and (ii) it makes it easy to emulate environments to show minimality results. 

We first define the moving source environment (MS) in which at every time at least one process (called the source) sends timely messages to all other processes, but this source may change over time and infinitely often. Although this environment is considerably weaker than a total synchronous environment, we show that it is still sufficient to implement registers, although it is not possible to implement the consensus abstraction. In fact, it can be emulated by hardware registers in totally asynchronous ``known'' networks for any number of process crashes. Therefore, if we would be able to implement consensus in this environment, we could contradict the famous FLP impossibility result \cite{214121}. This result states, that consensus cannot be implemented in asynchronous message passing networks, even if only one process may crash. Since we can emulate registers if only one process may crash \cite{200869}, we can also emulate the MS environment and therefore cannot be more powerful.

To implement consensus, we consider some additional stronger synchrony assumptions. Our first consensus algorithm assumes that additionally to the assumptions of the MS environment, eventually all processes communicate timely. We call this environment the eventual synchronous (ES) environment. It resembles Dwork et al.~\cite{Dwork88consensusin}. In our second consensus algorithm, we consider a weaker environment and only assume that eventually always the same process is able to send timely to all other processes. We call it the eventual stable source environment (ESS). It resembles the model of \cite{872081} in which it is used to elect a leader, a classical approach to implement in turn consensus.

Due to the indistinguishability of several processes that behave identical, a true leader election is not possible in our anonymous environment. Therefore, in our second algorithm, we take benefit of the fact that it suffices for the implementation of consensus if all processes that consider itself as a leader behave the same way. We show how to eventually guarantee this using the history of the processes proposal values.  

Furthermore, we consider the weak-set data-structure \cite{DF}. This data-structure comes along some problems that arise with registers in unknown and anonymous networks. Every process can add values to a weak-set and read the values written before. Contrary to a register, it allows for sharing information without knowing identities of other processes and without the risk of an overwritten value due to a concurrent write. Furthermore, we show that it precisely captures the power of the MS environment, i.e.~we can show that it can be implemented in the MS environment and a weak-set can be used to emulate the MS ennvironment. Interestingly, in known networks, a weak-set is equivalent to the register abstraction and can thus be seen as a generalization for unknown and anonymous networks.     

Furthermore, we show that although it is possible to emulate registers in our MS environment, it is not possible to emulate $\Sigma$ \cite{200377/IC}, the weakest failure detector for registers. And this result is not only due to the anonymity of the processes, it holds even if the number of processes and their identities are known. Note that this is not a contradiction, since the result in \cite{200377/IC} means only that $\Sigma$ is the weakest of all failure detectors with which a register can be implemented and we have exhibited synchrony assumptions where the existence of a failure detector is not necessary at all.

\subsection{Related work}

There have been several approaches to solve fault-tolerant consensus in anonymous networks deterministically. In \cite{DF}, fault-tolerant consensus is solved under the assumption that failure detector $\Omega$ \cite{journals/jacm/ChandraHT96} exists, i.e.~exactly one correct process eventually knows forever that it is the leader.    
In \cite{DBLP:journals/dc/GuerraouiR07}, fault-tolerant and obstruction-free\footnote{For obstruction-free consensus, termination is only guaranteed if a process can take enough steps without beeing interrupted by other processes.} consensus is solved if registers are available.

There has also been some research on systems where IDs are known but the number of processes is not. In \cite{1253054}, it is assumed that processes may crash, but furthermore that it is possible to detect the participants initially. In \cite{1432337}, a leader election algorithm for a system where infinitely many processes may join the system is presented if the number of processes simultaneously up is bounded.

To the best of our knowledge, this paper presents completely new approaches to emulate registers and solve the consensus problem in unknown and anonymous environments with partial synchrony.  


\section{Model and Definitions}

We assume a network with an unknown (but finite) number of processes where the processes have no IDs (i.e.~they are totally anonymous) and communicate using a broadcast primitive. The set of processes is denoted $\Pi$. We assume that the broadcast primitive is reliable, although it may not always deliver messages on time. Furthermore, any number of processes may crash and the processes do not recover. Processes that do not crash are called correct.

We model an algorithm $A$ as a set of deterministic automata, one for every process in the system. We assume only fair runs, i.e. every correct process executes infinitely many steps.    

\subsection{Consensus}
 
In the consensus problem, the processes try to decide on one of some proposed values. Three properties have to be satisfied:
\begin{description}[\IEEEsetlabelwidth{Termination:}] 
 \item[Validity:] Every decided value has to be a proposed value.
 \item[Termination:] Eventually, every correct process decides.
 \item[Agreement:] No two processes decide different values.
\end{description}

\subsection{An extension to GIRAF}

Algorithm \ref{alg:eGIRAF} presents an extension to the generic round-based algorithm framework of \cite{1449454} (GIRAF). It is extended to deal with the particularities of our model, namely the anonymity and unknown number of the processes. The framework is modeled as an I/O automaton. To implement a specific algorithm, the framework is instantiated with two functions: \initialize$()$ and \compute$()$. The $\compute()$ function takes the round number and the messages received so far as parameters. We omit to specify a failure detector output as parameter (as in \cite{1449454}), because we are not interested in failure detectors here. Both functions are non-blocking, i.e.~they are not allowed to wait for any 
other event. 

Our extension lies in the way we model the received messages. Since the processes have no IDs, we represent the messages that are received during one round as a set instead of an array.

The communication between the processes proceeds in rounds and the advancement of the rounds is controlled by the environment via the $\receive_i$ and $\endofround_i$ input actions. These actions may occur separately at each process $p_i$ and therefore rounds are not necessarily synchronized among processes. The framework can capture any asynchronous message passing algorithm (see \cite{1449454}). 

Environments are specified using round-based properties, restricting the message arrivals in each round. 

\begin{algorithm}[htb] 
  {\small \SetVline   
	\SetKwFor{mStates}{States:}{}{}
	\SetKwFor{mEffect}{effect:}{}{}
    \SetKwSwitch{sActions}{mInput}{mOutput}{Actions and Transitions:}{}{input}{output}{}
	       
\mStates{}{
	$k_i \in \mathbb{N}$, initially $0$; \\ 
	$M_i[\mathbb{N}] \subseteq \textrm{Messages}$, initially $\forall k \in \mathbb{N}: M_i[k] = \emptyset$\; 
}
		
\sActions{}{
    
	\mInput{$\endofround_i$}{
		\eIf{$(k_i=0)$}{
				$m := \initialize()$\;
		 }{
				$m := \compute(k_i, M_i)$\;
		}
		$M_i[k_i + 1] := M_i[k_i + 1] \cup \{ m \}$\;
		$k_i := k_i + 1$\;
			
    }
	\textbf{output} $\send(\langle M_i[k_i], k_i \rangle)_i$\;
	
	\mInput{$\receive(\langle M,k \rangle)_i$}{
		$M_i[k] := M_i[k] \cup M$\;
	}
	
}

\caption{Extended GIRAF generic algorithm for process $p_i$.}
\label{alg:eGIRAF}
}
\end{algorithm}

\subsection{Environments}

We say that a process $p_i$ is in round $k$, if there have been $k$ invocations of $\endofround_i$. A process $p_i$ has a \emph{timely link in round $k$}, if $\endofround_i$ occurs in round $k$ and every correct process $p_j$ receives the round $k$ message of $p_i$ in round $k$. 

In this paper, we consider three different environments: 
\begin{itemize}
 \item In the first one, which we call the moving-source (MS) environment, we assume that in every round $k$, there exists a process $p_s$ (a source) that has a timely link in round $k$.

 \item In the second environment, which we call the eventual synchronous (ES) environment, we demand the same as in the MS environment, but additionally require that there is some round $k$ such that in every round $k' \ge k$, all correct processes have timely links in round $k'$.
  
 \item In the third environment, which we call the eventually stable source (ESS) environment, we demand the same as in the MS environment, but additionally require that eventually the source process $p_s$ is always the same in every round. This means, that there is some round $k$ such that in every round $k' \ge k$, the same process $p_s$ has a timely link in round $k'$.
\end{itemize}

\section{Implementing consensus in ES} \label{sec:Cons}

\begin{algorithm}[htb] 
       {\small
\SetKwFor{On}{on}{do}{end}
\SetKw{WaitUntil}{wait until}       
\SetKwFunction{Compute}{$\compute$}       
\SetVline
       
\On{initialization}{
	$\val :=$ initial value\;
	$\written := \wlr := \proposed := \emptyset$\;
	\Return{$\proposed$}\; 
}
	
\On{$\compute(k_i, M_i)$}{
	$\written :=  \bigcap_{m \in M_i[k_i]} m$\;
	$\proposed := (\bigcup_{m \in M_i[k_i]} m) \cup \proposed$\; 
	\If{$(k_i \mod 2 = 0)$}{
		\uIf{$(\proposed = \wlr = \{\val\})$}{ 
			\textbf{decide} \val; \textbf{halt} 
		}	
		\ElseIf{$(\written \not= \emptyset)$}{
			$\val := \max(\written)$\;
		}	
		$\proposed := \{\val\}$\;
	}
	$\wlr := \written$\;
	\Return $\proposed$\;
}

\caption{A consensus algorithm in ES for process $p_i$.}
\label{alg:consensusSynch}
}
\end{algorithm}

Algorithm \ref{alg:consensusSynch} implements consensus in the ES environment. The idea of the algorithm is to ensure safety by waiting until a value is contained in every message received in a round. In this way, one can ensure that a value has also been relayed by the current source and is therefore known by everybody (we say that the value is written). If a process evaluates Line 9 to true, then $\val$ is known by everybody (because it was written in the last round) and no other process will consider another value as written, because only a value which has also been relayed by a source can be in $\written$. But the relayed value of a source would also be in $\proposed$ at every process. 

To guarantee the liveness of the consensus algorithm, we use the fact that eventually, all proposal values in the system are received in every even round by everybody and everybody will select the same maximum in Line 12. Therefore, everybody will propose the same value in the next round and the algorithm will terminate.

\subsection{Analysis}

For all local variables \textsc{var}, we denote by $\textsc{var}_i$ the local variable of process $p_i$ (e.g., $\proposed_i$). For every variable $\textsc{var}_i$, $\textsc{var}_i^k$ is the value of this variable after process $p_i$ has executed Line 7 when $\compute$ has been invoked with parameter $k$ (i.e.~in round $k$). 

\begin{lemma}
 \label{lem:nootherwrites}
 If no process has decided yet and for some $p_i$, $v \in \written_i^k$, then every process $p_j$ that enters round $k$ has $v \in \proposed_j^k$.
 \begin{IEEEproof}  
  If a process $p_i$ has a value $v$ in $\written_i^k$, then $v$ has been contained in every message, which $p_i$ has received in round $k$ (Line 6). This includes the message of the source, since by assumption the source has not yet terminated. But by definition, every other process $p_j$ that enters round $k$ also has received the message of this source in this round and added it to its set $\proposed_j^k$ (Line 7). Therefore, $v$ is in $\proposed_j^k$.  
 \end{IEEEproof}
\end{lemma}

\begin{lemma}
 \label{lem:writeeverywhere}
 If no process has decided yet and $p_i$ has $v \in \wlr_i^k$ in an even round $k$, then every other process $p_j$ that enters round $k$ has $v \in \written_j^k$.
 \begin{IEEEproof}
 If a process $p_i$ has a value $v$ in $\wlr_i^k$, then it has had $v$ in $\written_i^{k-1}$. Therefore, every other process $p_j$ that enters round $k-1$ has $v$ in $\proposed_j^{k-1}$ in the same odd round $k-1$ (Lemma \ref{lem:nootherwrites}). Since no value is removed from a set $\proposed$ in odd rounds, $v$ will be contained in every set $\proposed$ broadcast at the end of round $k-1$ and therefore get into $\written_j^{k}$ at every process $p_j$ that enters round $k$.   
 \end{IEEEproof}
\end{lemma}

\begin{theorem}
Algorithm \ref{alg:consensusSynch} implements consensus in the ES environment.
\begin{IEEEproof} 
We have to prove the 3 properties of consensus. Validity is immediately clear, because $\val$ is always an initial value.

To prove termination, assume that the system has stabilized, i.e.~all faulty processes have crashed and all messages are received in the round after which they have been sent. Then, all processes receive the same set of messages in every round. Therefore, the set $\proposed$ and thus $\written$ is the same at all correct processes and everybody will always select the same maximum in Line 12. In the next round all processes start with the same proposal value and this value will be written in every future round. Thus, everybody will evaluate Line 9 to true in the next round.    

To prove agreement, assume $p_i$ is the first process that decides a value $v$ in a round $k$. This means, that $p_i$ has evaluated Line 9 to true. If some other value than $v$ would have been written anywhere in the system, this would contradict $\proposed = \{v\}$ (Lemma \ref{lem:nootherwrites}), since $p_i$ is the first process that decides. Furthermore, $v$ is in $\written$ at every process in the system in round $k$, since it is also in $\wlr$ (Lemma \ref{lem:writeeverywhere}). Therefore, every other process decides $v$ in the same round, or it will evaluate Line 11 to true and select $v$ as new $\val$. Thus, no other value will ever get into $\proposed$ anywhere in the system, no other value will ever be written and no other value will ever be selected as $\val$.

\end{IEEEproof}
\end{theorem}

\section{Implementing consensus in ESS} \label{sec:ConsESS}

Algorithm \ref{alg:consensusESS} implements consensus in the ESS environment. For the safety part, the algorithm is very close to algorithm \ref{alg:consensusSynch} (see Section \ref{sec:Cons}). 

To guarantee liveness, we use the fact that we have at least one process which is eventually a source forever. We use the idea of the construction of the leader failure detector $\Omega$ \cite{journals/jacm/ChandraHT96}. It elects a leader among the processes which is eventually stable. In ``known'' networks, with some eventual synchrony, $\Omega$ can be implemented by counting heartbeats of processes (e.g.~in \cite{872081}). But we are not able to count heartbeats of different processes here, because in our model the processes have no IDs. To circumvent this problem, we identify processes with the history of their proposal values. If several processes have the same history, they either propose the same value, or their histories diverge and will never become identical again. Eventually, all processes will select the same history as maximal history and the processes with this history will propose in every round the same values. 

\subsection{Implementation}

Every process maintains a list of the values it broadcasts in every round (specifically, its proposal values). This list is denoted by the variable $\history$. In this way, two processes that propose in the same round different values will eventually have different $\history$ variables. Note that, although the space required by the variables may be unbounded, in every round they require only finite space. Thus, if we could ensure that eventually all processes that propose have in every round the same history (and at least one process proposes infinitely often), then the proposal values sent are indistinguishable from the proposal values of a single ``classical'' leader.

However, the history of a process permanently grows. Therefore, every process includes its current history in every message it broadcasts. Furthermore, it maintains a counter $\C$ for every history it has yet heard of (in such a way that no memory is allocated for histories it has not yet heard of). Then, it compares the histories it receives with the ones it has received in previous rounds. If some old history is a prefix of a new history, it assigns the counter of the new history the value of the counter of the old one, increased by one. Thus, the counter of a history that corresponds to an eventual source is eventually increased in every round. 

In this way, it is possible to ensure that eventually only eventual sources that converge to the same infinite history consider itself as leader. In a classical approach, eventually only these leaders would propose values. But to meet our safety requirements, it is crucial to ensure that all processes propose in every round at least something to make sure that the value of the current source is received by everybody. Therefore, we let processes that do not consider itself as a leader propose the special value $\bot$. 

\begin{algorithm}[htb] 
       {\small
\SetKwFor{On}{on}{do}{end}
\SetKw{WaitUntil}{wait until}       
\SetKwFunction{Compute}{$\compute$}       
\SetVline
       
\On{initialization}{
	$\val :=$ initial value; $\forall \h, \C[\h] := 0$; $\history := \val$\;
	$\written := \wlr := \proposed := \emptyset$\;
	\Return{$\m = \langle \proposed,\history,\C \rangle$}\;
}
	
\On{$\compute(k_i, M_i)$}{
	\nllabel{l:written} $\written :=  \bigcap_{m \in M_i[k_i]} m.\proposed$\;
	\nllabel{l:proposed} $\proposed := (\bigcup\limits_{m \in M_i[k_i]} m.\proposed) \cup \proposed$\;             
	$\forall \h, \C[\h] := \min_{m \in M_i[k_i]}(m.\C[\h])$\;
	\nllabel{l:min} $\forall m \in M_i[k_i], \C[m.\history]$ := $1$ $+$ $\max \{\ C[\h] \mid \h \textrm{ is a prefix of } m.\history \}$\;
            
	\nllabel{l:mod} \If{$(k_i \mod 2 = 0)$}{ 
		\nllabel{l:decision} \uIf{$(\wlr = \{\val\}) \wedge (\proposed \subseteq \{\val, \bot\})$}{
			\textbf{decide} \val; \textbf{halt}\;
		}	
		\nllabel{l:select} \ElseIf{$(\written \setminus \{\bot\} \not= \emptyset)$}{
			\nllabel{l:max} $\val := \max(\written \setminus \{\bot\})$\;
		}
		\nllabel{l:proposal} \eIf{($\forall \h, \C[\history] \ge \C[\h]) \vee (\proposed \subseteq \{\val, \bot\}$)}{
			\nllabel{l:proposalselect} $\proposed := \{\val\}$\;
		}{$\proposed := \{\bot\}$\;}
	}
	
	$\wlr := \written$\;
	$\written := \proposed$\;
	append $\val$ to $\history$\;
	\Return{$\m = \langle \proposed,\history,\C \rangle$}\;
}

\caption{The consensus algorithm in ESS for process $p_i$.}
\label{alg:consensusESS}
}
\end{algorithm}

\subsection{Analysis}
Similarly to Section \ref{sec:Cons}, for every variable $\textsc{var}_i$, $\textsc{var}_i^k$ is the value of this variable after process $p_i$ has executed Line 9 in round $k$. 
\begin{definition}
 We say, that $p_i$ \emph{has heard of} $p_j$'s round $k$ message ($\m_{j}^k$), if $p_i$ has received $\m_{j}^k$ in round $k$, or if there exists another process $p_l$ such that $p_i$ has heard of $p_l$'s round $k'$ message for some $k' > k$ and $p_l$ has heard of $p_j$'s round $k$ message.  
\end{definition}
Let process $p_s$ be an eventual source. We then identify three groups of processes:
\begin{description}[\IEEEsetlabelwidth{$\out$:}] 
 \item[$\out$:] The processes, the eventual source $p_s$ has infinitely often heard of.
 \item[\textit{$\diamond$-silent}:] The processes that are not $\out$.
 \item[$\proposer$:] The $\out$ processes that have eventually in every round timely links towards all other $\out$ processes.\footnote{Note that it is possible that the message an $\out$ process actually has received is not the message that a $\proposer$ has sent. It is sufficient if it receives an identical message from another process.}
 \item[$\leader$:] We say that a process $p_i$ is a \emph{leader} in some round $k$ ($p_i \in \leader(k)$), iff $ \forall \h, \C_{i}^k [\history_{i}^k] \ge \C_{i}^k[\h] $. 

 If process $p_i$ is eventually a leader forever, i.e.~there exists a $k$, such that for all $k' \ge k$, $p_i \in \leader(k')$, then we simply write that $p_i \in leader$. Note that it may be possible that there are several processes in $\leader$.
\end{description}
The sets relate to each other in the following way:
\begin{eqnarray*}
 & \{ p_s \} \subseteq \proposer \subseteq \out \subseteq \textit{correct} & \\
 & \textrm{and } \ \textit{$\diamond$-silent} \cap \out = \emptyset &
\end{eqnarray*}
We will later show that $\leader \subseteq \proposer$ (Lemma \ref{lem:propmaximal}).


\begin{lemma} \label{lem:propwritten}
 Eventually, in every odd round $k$, for every $\proposer$ $p_i$, the set $\proposed$ in $\m_{i}^k$ is a subset of the set $\written$ at all $\out$ processes in round $k+1$. More formally:
 \begin{eqnarray*}
  & \exists k, \forall k' \ge k \mbox{ with } k' \emph{ mod } 2 = 1, &  \\
  & \forall p_i \in \proposer, \forall p_j \in \out: & \\
  & \m_{i}^{k'} = \langle \proposed, -, - \rangle & \\
  & \rightarrow \proposed \subseteq \written_{j}^{k'+1} & \\
 \end{eqnarray*}
 \begin{IEEEproof}
  Follows directly from the definition of $\diamond$-proposers and the fact that out-connected processes eventually do not receive any timely messages from $\diamond$-silent processes.
 \end{IEEEproof}

\end{lemma}

\begin{lemma}
 \label{lem:by1}
 Eventually, at all $\out$ processes, the counters that correspond to histories of $\proposer$s increase in every round by one. More formally:
 \begin{eqnarray*}
  \exists k, \forall k' \ge k, \forall p_i \in \proposer, \forall p_j \in \out, \\ 
  \C_{j}^{k'+1} [\history_{i}^{k'+1}] = \C_{j}^{k'} [\history_{i}^{k'}] + 1 
 \end{eqnarray*}
 \begin{IEEEproof}
  Assume a time when the system has stabilized. This means, that all $\proposer$s send timely messages to all $\out$ processes in every round and no $\out$ process receives timely messages from $\silent$ processes. Then, let $k$ be the number of the current round and for every $\proposer$ $p_i$ let $p_j$ be an $\out$ process, such that the counter $\C_{j}^k [\history_{i}^k]$ is minimal among all $\out$ processes in round $k$. Then, the counter for $p_i$'s history at $p_j$ will never decrease, because $p_j$ will never receive a message with a lower counter from any other process.
  
Since $p_i$ is a $\proposer$, the counter for $p_i$'s history will increase by one at $p_j$ in every round. For every other $\out$ process, since it receives also a message from $p_i$ in every round and it can only finitely often receive a lower counter corresponding to $p_i$'s history (the lowest one is $p_j$'s), the counter of $p_i$'s history eventually increases in every round by one. 
 \end{IEEEproof}

\end{lemma}

\begin{lemma} \label{lem:infhigher}
If a history of a process $p_j$ infinitely often corresponds to a maximal counter at a $\proposer$ $p_i$, then $p_j$ is a leader forever. More formally:
 \begin{eqnarray*}
  & \forall p_i \in \proposer, \forall p_j \in \Pi: & \\
  & (\forall k, \exists k' > k, \forall h, 
       (\C_{i}^{k'} [\history_{j}^{k'}] \ge \C_{i}^{k'} [h]) ) & \\
  & \rightarrow \ p_j \in \leader &
 \end{eqnarray*}
 
 \begin{IEEEproof}
 We first show that $p_j \in \proposer$. Assume that it is not. Since $p_i \in \proposer$, eventually the counter that corresponds to $p_i$'s history is increased by one at every $\out$ process (Lemma \ref{lem:by1}). Since $p_j \not\in \proposer$, some $\out$ process $p_l$ does not receive $\m_{j}^k$ in round $k$ for infinitely many rounds $k$. Therefore, the counter at $p_l$ that corresponds to $p_j$'s history is not increased by one in these rounds and is eventually  strictly lower than the one that corresponds to $p_i$'s history. Since every time some $\out$ process has a lower counter than the others, eventually this counter propagates to all other $\out$ processes, $p_i$'s history will eventually be higher than $p_j$'s at all $\out$ processes. A contradiction.
  
 If $p_i$ and $p_j$ are both $\proposer$s, then eventually they receive their messages timely in every round $k$. Since $p_j$'s history increases at all $\out$ processes by one (Lemma \ref{lem:by1}), eventually $\C_{j}^{k} [\history_{j}^{k}] = \C_{i}^{k} [\history_{j}^{k}]$. Since by our assumption, in some future round $k'$, $p_j$'s history is maximal at $p_i$ and a counter can increase by at most one and the counters that correspond to $p_j$'s history increase always by one (Lemma \ref{lem:by1}), $\C_{j}^{k} [\history_{j}^{k}]$ is maximal forever and therefore $p_j$ is a leader forever.
 \end{IEEEproof}
\end{lemma}

\begin{lemma} \label{lem:propmaximal}
Eventually, there exists a process $p_i \in \leader$ and every leader is a $\proposer$. More formally:
 \begin{eqnarray}
  \exists k, \exists p_i \in \Pi, \forall k' \ge k: p_i \in \leader(k') \label{lem:b} \\
  \textrm{and } \  \forall p_i \in \Pi: (\forall k, \exists k', k' > k, p_i \in \leader(k'))  \nonumber \\  
 \rightarrow \ p_i \in \proposer \label{lem:a}
 \end{eqnarray}
\begin{IEEEproof}
The eventual source $p_s$ is a $\proposer$. Therefore, there exists at least one $\proposer$. Either $p_s$ is also a leader forever, or there is another process whose history infinitely often corresponds to a higher counter at $p_s$ than $p_s$'s history. Then, with Lemma \ref{lem:infhigher} this process is a leader forever. This implies (\ref{lem:b}).

Assume a process $p_i$ is not a $\proposer$. Then, $p_i$'s counter is increased by less than one in infinitely many rounds at some processes. Because eventually these counters propagate to all $\out$  processes and the values of $\proposer$s are increased in every round by at least one (Lemma \ref{lem:by1}), eventually the history of some $\proposer$ is higher than that of $p_i$. Therefore, $p_i$ cannot be a leader forever. This implies (\ref{lem:a}).

\end{IEEEproof}
\end{lemma}


\begin{lemma} \label{lem:nodifferent}
If no process has decided yet, then eventually only values of leaders and $\bot$ get into a set $\written$ anywhere. More formally:
\begin{eqnarray*}
 &\exists k, \forall k' \ge k, \forall p_i \in \Pi: & \\
 & \written_{i}^{k'} \subseteq \cup_{p_j \in \leader(k')} \val_{j}^{k'} \cup \{\bot\} &
\end{eqnarray*}
 \begin{IEEEproof}
 There is a time after which there exists at least one leader and all leaders are $\proposer$s (Lemma \ref{lem:propmaximal}) and since leaders propose their values always, all their values get into every set $\written$ at all out-connected processes in every even round (Lemma \ref{lem:propwritten}).
 
 Therefore, every set $\proposed$ contains a value of a leader (compare Lemma \ref{lem:nootherwrites}) and no process that considers itself not as leader and has a value different from a leader will evaluate line 15 to true and add a different value to its set $\proposed$.
 \end{IEEEproof} 
\end{lemma}

\begin{theorem}
Algorithm \ref{alg:consensusESS} implements consensus in ESS.
\begin{IEEEproof} 
 We have to prove the 3 properties of consensus. Validity is clear, since $\val$ is always an initial value.
 
 To prove termination, assume there exists a run where no process ever decides. Then, eventually only non-$\bot$ values of leaders will get into a set $\written$ anywhere (Lemma \ref{lem:nodifferent}) and they will get into $\written$ always in every even round (Lemma \ref{lem:propwritten}) and all $\out$ processes select the same value (the maximum in Line 14). Therefore, only this value and $\bot$ will be written in subsequent rounds and every $\out$ process will select this value as value for $\proposed$ in Line 16 (i.e., no $\out$ process will select $\bot$) and everybody will evaluate Line 11 to true in the next round. Therefore, eventually, every correct process will decide.    
 
 To prove agreement, assume $p_i$ is the first process that decides a value $v$ in a round $k$. This means, that $p_i$ has evaluated Line 11 to true. Then, as $\proposed \subseteq \{v, \bot\}$, no other value different from $\bot$ is in a set $\written$ anywhere in the system (compare Lemma \ref{lem:nootherwrites}) and $v$ is in $\written$ at every process in the system in round $k$, since it is also in $\wlr$ (compare Lemma \ref{lem:writeeverywhere}). Therefore, every other process decides $v$ in the same round, or it will evaluate Line 13 to true and select $v$ as new $\val$ and no other value different from $\bot$ will ever get into $\proposed$ anywhere in the system and therefore, no other value will ever be selected as $\val$.

\end{IEEEproof}
\end{theorem}

\section{Weak-Sets}

The weak-set data structure has been introduced by Delporte-Gallet and Fauconnier in \cite{DF}.

A weak-set $S$ is a shared data structure that contains a set of values. It is defined by two operations: the $\add_S(v)$ operation to add a value $v$ to the set and the $\get_S$ operation which returns a subset of the values contained in the weak-set. Note that we do not consider operations to remove values from the set. Every $\get_S$ operation returns all values $v$ where the corresponding $\add_S(v)$ operation has completed before the beginning of the $\get_S$ operation. Furthermore, no value $v'$ where no $\add_S(v')$ has started before the termination of the $\get_S$ operation is returned. For $\add_S$ operations concurrent with the $\get_S$ operation, it may or may not return the values. Therefore, weak-sets are not necessarily linearizable\footnote{A weak object is linearizable (also called atomic) if all of its operations appear to take effect instantaneously \cite{67423}.}.

\subsection{Weak-Sets and registers} 

A weak-set is clearly stronger than a (regular) register:
\begin{proposition}
  A weak-set implements a (regular) multiple-writer multiple-reader register.
 \begin{IEEEproof}
  To write a value, every process reads the weak-set and stores the content in a variable \textsc{history}. Then, every process adds the value to be written together with \textsc{history} to the weak-set.
  
  To read a value, a process reads the weak-set and returns the highest value among all values accompanied by a \textsc{history} with maximal length.
  
  This transformation satisfies the two properties of regular registers, namely termination and validity. Termination follows directly from the termination property of weak-sets.
  
  If several processes write at the same time, two reads at two different processes may return different values, but after all writes have completed, the return value will be the same at all processes. To see that also validity holds, consider the value returned by a read. If there is no concurrent write, then the value returned is the last value written (i.e.~the maximal value of all values concurrently written). 
  
 \end{IEEEproof} 
\end{proposition}

In \cite{DF}, a weak-set is implemented using (atomic) registers in the following two cases:

\begin{proposition} \label{prop:WSbyID}
 If the set of processes using the weak set is known (i.e.~the IDs and the quantity), then weak-sets can be implemented with single-writer multiple-reader registers.
 
\end{proposition}

\begin{proposition} 
 If the set of possible values for the weak set is finite, then weak-sets can be implemented with multiple-writer multiple-reader registers.
 
\end{proposition}

\subsection{Weak-Sets and the MS environment} 
 
Algorithm \ref{alg:WeakSet} shows how to implement a weak-set in the MS environment.
Similarly to Section \ref{sec:Cons}, for every variable $\textsc{var}_i$, $\textsc{var}_i^k$ is the value of this variable after process $p_i$ has executed Line 15 in round $k$ (i.e.~after $\compute$ is called with parameter $k$). 

\begin{algorithm}[htb] 
{\small       
\SetKwFor{On}{on}{do}{end}
\SetKw{WaitUntil}{wait until}       
\SetKwFunction{Compute}{$\compute$}       
\SetVline
       
\On{initialization}{
	$\val := \bot$; 	$\proposed := \written := \emptyset$\;
	$\textsc{block} := \qfalse$\;
	\Return{$\proposed$}\;
}

\On{get}{
	 \Return{$\proposed$}\;
}

\On{add$(v)$}{
	$\proposed := \proposed \cup \{v\}$\;
	$\val := v$\;
	$\textsc{block} := \qtrue$\;
	\WaitUntil{$(\textsc{block} = \qfalse)$}\;
	\Return{ack}\;
}
	
\On{$\compute(k_i, M_i)$}{
	$\written :=  \bigcap_{m \in M_i[k_i]} m$\;
	$\proposed := (\bigcup_{m \in M_i[k'], 1 \le k' \le k_i} m) \cup \proposed$\; 
	\lIf{$(\val \in \written)$}{
		$\textsc{block} := \qfalse$\;
	}	
	\Return{$\proposed$}\;
}

\caption{A weak-set algorithm in the MS environment for process $p_i$.}
\label{alg:WeakSet}
}
\end{algorithm}

\begin{lemma}
 \label{lem:safetyWeakSet}
 If for some $p_i$, $v \in \written_i^k$, then every process $p_j$ that enters round $k$ has $v \in \proposed_j^k$.
 \begin{IEEEproof} 
 The proof is analogous to Lemma \ref{lem:nootherwrites}.
 \end{IEEEproof}
\end{lemma}

\begin{lemma}
 \label{lem:writtenforever}
 If some value is in $\written$ at some process, then this value will be forever in $\proposed$ at all processes.
 \begin{IEEEproof} 
 Since it is never a value removed from any set $\proposed$, this follows immediately from Lemma \ref{lem:safetyWeakSet}.
 \end{IEEEproof}
\end{lemma}

\begin{theorem}
 Algorithm \ref{alg:WeakSet} implements a weak-set. 
 \begin{IEEEproof}
 We have to show that all operations terminate at all correct processes and that every $\get$ operation returns all values which have been added before.  
 
 The only position where an operation may be blocked is in Line 11. But since eventually all messages will be received by all correct processes, every value will eventually be in every set $\proposed$ and therefore eventually be in every set $\written$. Thus, no correct process will block in Line 11 forever.
 
 To show that every $\get$ operation returns all values which have been added before, see that an $\add(v)$ operation only terminates if $v$ is in $\written$ at some process. Together with Lemma \ref{lem:writtenforever}, this means that this value will be returned by every process in Line 6.    

 \end{IEEEproof}
\end{theorem}

\subsection{Emulation of the MS environment with weak-sets} 

Algorithm \ref{alg:MS} emulates the MS environment using a weak-set $S$ and the corresponding $\add_S$ and $\get_S$ methods.

As a weak-set is implementable by only using registers (see Proposition \ref{prop:WSbyID}) and the FLP impossibility result \cite{214121} states that consensus is not implementable using only registers, this implies, that it is not possible to implement consensus in the MS environment (without any additional assumptions like in ES).

\begin{algorithm}[htb] 
{\small       
\SetKwFor{On}{on}{do}{end}
\SetKw{Trigger}{trigger}       
\SetVline
       
\On{initialization}{
         $\delivered := \emptyset$\;
         \Trigger{$\endofround_i$}\;
}
\BlankLine

\On{$\send(m_i,k_i)_i$}{
	$\add_S(\langle m_i,k_i \rangle)$\;
	\ForAll{$\langle m,k \rangle \in \get_S \setminus \delivered$}{
		$\delivered := \delivered \cup \{\langle m,k \rangle\}$\;
		\Trigger{$\receive(m,k)_i$}\;
	}	
	\Trigger{$\endofround_i$}\;
}

\caption{Emulating the MS environment for process $p_i$ using a weak-set $S$.} 
\label{alg:MS}
}
\end{algorithm}

\begin{theorem}
 Algorithm \ref{alg:MS} emulates the MS environment. 
 \begin{IEEEproof}
 Clearly, eventually all messages get delivered and all correct processes execute an infinite number of rounds.
 
It remains to show, that in every round $k$, there exists a process $s_k$ such that for every process $p_i$ at which $\endofround_i$ occurs in round $k$, $p_i$ receives the round $k$ message of $s_k$ in round $k$.

Let $p_i$ be the first process that finishes to add the value of a round $k$. If several processes finish to add their values at exactly the same time, choose one. 

\textbf{Claim:} Every process at which $\endofround$ is triggered in round $k$ has  received $p_i$'s round $k$ value.

The proof is by contradiction. Assume that a process $p_j$ triggers $\endofround$ in round $k$ without having received $p_i$'s round $k$ value. By the definition of a weak-set, this means that $p_j$'s $\get_S$ begun before $p_i$'s $\add_S$ was completed. But a process will only start a $\get_S$ after it has finished to add its own value. A contradiction to the fact that $p_i$ was the first process that has completed its $\add_S$.   
 \end{IEEEproof}
\end{theorem}

\section{The MS-environment and the $\Sigma$ failure detector}

The quorum failure detector $\Sigma$ \cite{200377/IC} outputs lists of IDs of trusted processes (i.e.~it is not well-defined in our anonymous model) and it satisfies the following properties:
\begin{description}[\IEEEsetlabelwidth{Completeness:}] 
 \item[Intersection:] Given any two lists of trusted processes, possibly at different times and by different processes, at least one process belongs to both lists. 
 \item[Completeness:] Eventually at all correct processes, every trusted process is correct.
\end{description}

$\Sigma$ has been shown to be the weakest failure detector to emulate registers in totally asynchronous message-passing systems \cite{200377/IC} (with known IDs). This means, that $\Sigma$ is sufficient to emulate registers in such systems and with any failure detector which is also sufficient to implement registers in such a system, it is possible to emulate $\Sigma$. 
Interestingly, although it is possible to implement a register in the MS environment (via weak-sets), we show that even if we assume that the number of processes and their IDs are known, it is not possible to emulate $\Sigma$. Note that this is no contradiction, since in our model no failure detector is necessary for the emulation.

\begin{proposition}
 It is not possible to emulate $\Sigma$ in the MS-environment, even if the number of processes and their IDs are known.
 \begin{IEEEproof}
  Assume there exists such an algorithm and consider a run $r_1$ where process $p_1$ is the only correct process, $p_1$ is always the source, and $p_1$ receives no messages from other processes. Then, by the completeness property of $\Sigma$, there exists some time $t$ after which the output of $\Sigma$ is $\{p_1\}$.

  Similarly, consider a run $r_2$ where process $p_2$ is the only correct process and $p_1$ crashes after time $t$. Again, $p_1$ is the source until time $t$ and receives no messages from other processes (this is possible, since the messages from $p_2$ may be arbitrary delayed). For $p_1$, run $r_1$ and $r_2$ are indistinguishable up to time $t$ and consequently the $\Sigma$ at $p_1$ will output $\{p_1\}$ at $p_1$ at time $t$. But since eventually, the output at $p_2$ has to be $\{p_2\}$ forever, this contradicts to the intersection property of $\Sigma$.
 
 \end{IEEEproof}
\end{proposition}

\section{Conclusions}

This paper has provided algorithms to emulate registers and solve consensus under different synchrony assumptions in unknown and anonymous message-passing networks that suffer from crash-failures. One of these algorithms uses a novel pseudo leader election primitive. 


Furthermore, we have shown that the MS environment (i.e.~a system with a moving timely source) is equivalent to weak-sets, a generalization of registers for unknown and anonymous systems. In some sense, this indicates that the synchrony assumptions in this environment are necessary to implement basic safety primitives.

Additionally, we have shown that in the MS environment, it is not possible to emulate $\Sigma$, the weakest failure detector to emulate registers \cite{200377/IC}, even if we assume the existence of IDs and a bound on the number of processes. 
To the best of our knowledge, we found for the first time a partially synchronous environment in which registers are implementable and $\Sigma$ is not.



\IEEEtriggeratref{11} 

\bibliographystyle{plain}
\bibliography{bibtex}

\end{document}